\documentclass[showpacs,10pt,twocolumn,prl]{revtex4-1}

\usepackage{amsmath}
\usepackage{amssymb}
\usepackage{graphicx}
\usepackage{amssymb}
\usepackage{graphics}
\usepackage{epsfig}
\usepackage{CJK}
\usepackage{color}
\usepackage{soul}

\begin{document}

\begin{CJK*}{GBK}{Song}
\title{Magnetic critical behavior and anomalous Hall effect in 2H-Co$_{0.22}$TaS$_{2}$ single crystals}
\author{Yu Liu,$^{1,*}$ Zhixiang Hu,$^{1,2}$ Eli Stavitski,$^{3}$, Klaus Attenkofer,$^{3,\dag}$ and C. Petrovic$^{1,2}$}
\affiliation{$^{1}$Condensed Matter Physics and Materials Science Department, Brookhaven National Laboratory, Upton, New York 11973, USA\\
$^{2}$Materials Science and Chemical Engineering Department, Stony Brook University, Stony Brook, New York 11790, USA\\
$^{3}$National Synchrotron Light Source II, Brookhaven National Laboratory, Upton, New York 11973, USA}
\date{\today}

\begin{abstract}
  We report ferromagnetism in 2H-Co$_{0.22}$TaS$_2$ single crystals where Co atoms are intercalated in the van der Waals gap, and a systematic study of its magnetic critical behavior in the vicinity of $T_c \sim 28$ K. The obtained critical exponents $\beta$ = 0.43(2), $\gamma$ = 1.15(1), and $\delta = 3.54(1)$ fulfill the Widom scaling relation $\delta = 1+\gamma/\beta$ and follow the scaling equation. This indicates that the spin coupling in 2H-Co$_{0.22}$TaS$_2$ is of three-dimensional Hersenberg type coupled with long-range magnetic interaction, and that the exchange interaction decays with distance as $J(r)\approx r^{-4.69}$. 2H-Co$_{0.22}$TaS$_2$ exhibits a weak temperature-dependent metallic behavior in resistivity and negative values of thermopower with dominant electron-type carriers, in which obvious anomalies were observed below $T_c$ as well as the anomalous Hall effect (AHE). The linear scaling behavior between the modified anomalous Hall resistivity $\rho_{xy}/\mu_0H$ and longitudinal resistivity $\rho_{xx}^2M/\mu_0H$ implies that the origin of AHE in 2H-Co$_{0.22}$TaS$_2$ should be dominated by the extrinsic side-jump mechanism.
\end{abstract}
\maketitle
\end{CJK*}

\section{INTRODUCTION}

Recent discovery of intrinsic long-range magnetic order in a mono- or few-layer in two-dimensional (2D) van der Waals (vdW) magnets \cite{Geim,Novoselov,Lee,Huang,Gong,Deng,Bonilla,Hara}, such as FePS$_3$, Cr$_2$Ge$_2$Te$_6$, CrI$_3$, Fe$_3$GeTe$_2$, VSe$_2$ and MnSe$_2$, has provided new platforms for studying fundamental 2D magnetism and designing novel spin-related devices. Interestingly, the spin interactions in 2D vdW magnets exhibit different types of critical behavior, such as 2D Heisenberg type in MnPS$_3$, 2D XY type in NiPS$_3$, and 2D Ising type in FePS$_3$ \cite{Joy}. Three-dimensional (3D) Heisenberg behavior is observed in bulk CrI$_3$ and Fe$_3$GeTe$_2$ \cite{YuLiu1,BJLiu1,YuLiu2,GTLin1}, however, a crossover from 3D to 2D Ising ferromagnetism is revealed in thickness-reduced crystals \cite{YuLiu4,Fei}. 2D Ising behavior is confirmed in Cr$_2$(Si,Ge)$_2$Te$_6$ \cite{BJLiu0,YuLiu0,WLiu0,GTLin0}, whereas mean field type interaction is found in its 3D analog of Mn$_3$Si$_2$Te$_6$ since an additional 1/3 Mn residing in vdW gap induced stronger interlayer coupling \cite{YuLiu3}.

Intercalated transition metal dichalcogenides commonly feature 3$d$ atoms in vdW gap and exhibit diverse magnetic orders \cite{Parkin,Friend0}. For instance, 2H-M$_{0.33}$TaS$_2$ with M = V, Cr, Mn is ferromagnetic (FM), with M = Co, Ni is antiferromagnetic (AFM), whereas with M = Fe exhibits both types of magnetic behavior \cite{Parkin}. 2H-Mn$_x$TaS$_2$ is FM with easy $\mathbf{ab}$ plane \cite{Hinode,Zhang1}, 2H-Fe$_x$TaS$_2$ is FM with easy $\mathbf{c}$ axis for 0.2 $\leq$ $x$ $\leq$ 0.4 but is AFM for higher Fe content, then 2H-(Co,Ni)$_x$TaS$_2$ is AFM, because the super-exchange interactions become larger as the intercalate ion is varied from Mn through Fe and Co to Ni. Among these materials, the most interesting member is 2H-Fe$_{0.25}$TaS$_2$, the only member that exhibits FM order with strong uniaxial anisotropy, similar to recently investigated 2D vdW magnets. It features large magnetocrystalline anisotropy, magnetoresistance, and sharp switching in magnetization \cite{Morosan,Hardy,Chen}. Anomalous Hall effect (AHE) mechanism in 2H-Fe$_{0.29}$TaS$_2$ changes from extrinsic scattering to intrinsic contribution with thickness reduction \cite{Cai}, whereas the spin coupling in 2H-Fe$_{0.26}$TaS$_2$ is of 3D Heisenberg type coupled with long-range magnetic interactions \cite{Zhang}.

In this work, we synthesized 2H-Co$_{0.22}$TaS$_2$ single crystal that shows FM order with $T_c$ $\sim$ 28 K and a strong uniaxial anisotropy. Several methods were employed to determine the critical exponents so as to shed light on the mechanism of the paramagnetic (PM)-FM transition. Critical exponents $\beta = 0.43(2)$ with $T_c$ = 28.55(2) K and $\gamma = 1.15(1)$ with $T_c$ = 28.62(2) K are obtained by the Kouvel-Fisher (KF) method, whereas $\delta = 3.54(1)$ is obtained by critical isotherm at $T_c$. The renormalization group theory analysis suggests that the spin coupling in 2H-Co$_{0.22}$TaS$_2$ is of 3D Heisenberg type coupled with long-range magnetic interaction; the exchange interaction decays with distance as $J(r)\approx r^{-4.69}$. 2H-Co$_{0.22}$TaS$_2$ shows weak temperature-dependent metallic behavior in resistivity $\rho(T)$ and negative values of thermopower $S(T)$ with dominant electron-type carriers as well as the AHE below $T_c$. A linear scaling behavior between the modified anomalous Hall resistivity $\rho_{xy}/\mu_0H$ and longitudinal resistivity $\rho_{xx}^2M/\mu_0H$ indicates the extrinsic side-jump mechanism of AHE.

\section{EXPERIMENTAL DETAILS}

\begin{figure*}
\centerline{\includegraphics[scale=1]{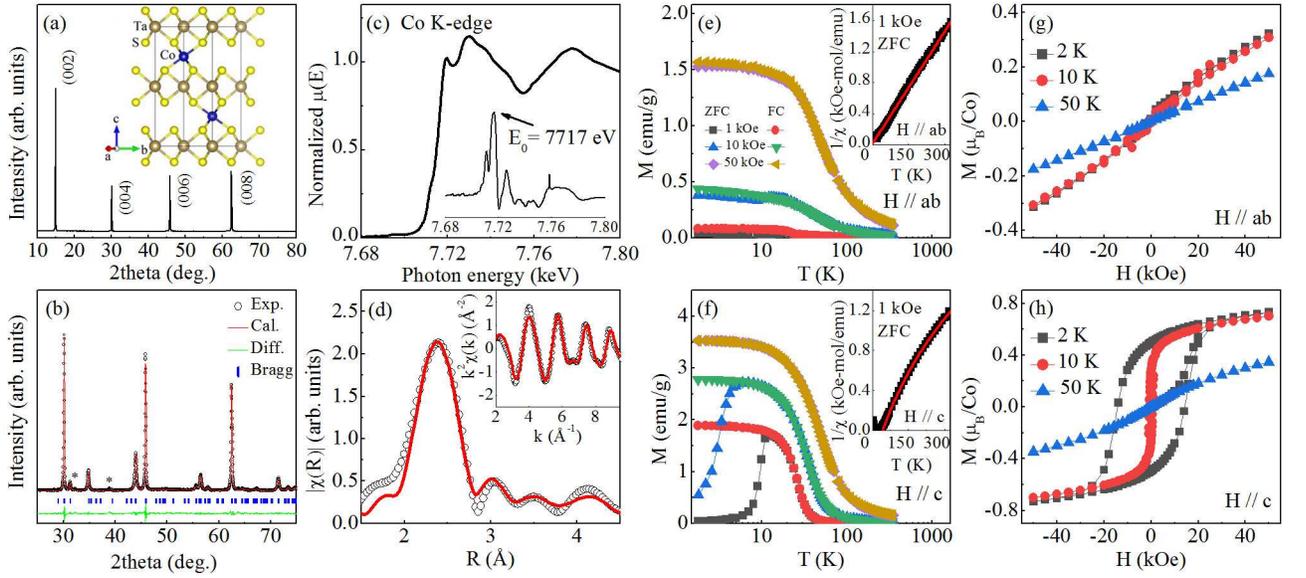}}
\caption{(Color online) (a) Single crystal and (b) powder x-ray diffraction (XRD) patterns of 2H-Co$_{0.22}$TaS$_2$. Inset in (a) shows the crystal structure of 2H-Co$_{0.22}$TaS$_2$. The tiny extra peaks [labeled as asterisk in (b)] at 32.2$^\circ$ and 38.9$^\circ$ are from pure iodine and sulfur, respectively. (c) Normalized Co $K$-edge x-ray absorption near edge structure (XANES) spectra. Inset shows the derivative $\mu(E)$ curve. (d) Fourier transform magnitudes of the extended x-ray absorption fine structure (EXAFS) oscillations (symbols) for Co $K$-edge with the phase shift correction. The model fits are shown as solid lines. Inset shows the corresponding filtered EXAFS (symbols) with $k$-space model fits (solid lines). Temperature-dependent magnetization $M(T)$ with both zero-field cooling (ZFC) and field cooling (FC) modes for (e) $\mathbf{H} \parallel \mathbf{ab}$-plane and (f) $\mathbf{H} \parallel \mathbf{c}$-axis at various magnetic fields. Insets show the corresponding inverse susceptibility with H = 1 kOe fitted by the Curie-Weiss law (solid lines). Field-dependent magnetization for (g) $\mathbf{H} \parallel \mathbf{ab}$-plane and (h) $\mathbf{H} \parallel \mathbf{c}$-axis at various temperatures.}
\label{XRD}
\end{figure*}

Single crystals of 2H-Co$_{0.22}$TaS$_2$ were grown by chemical vapor transport method with iodine agent. The raw materials (Co, Ta, and S powders) were mixed with a molar ratio of 0.25 $:$ 1 $:$ 2, sealed in an evacuated quartz tube, and then heated for two weeks in a two-zone furnace with the source zone temperature of 1000 $^\circ$C and the growth zone temperature of 900 $^\circ$C, respectively. The obtained single crystals are hexagonal shape with typical dimensions as $4\times4\times0.5$ mm$^3$. The average stoichiometry was determined by examination of multiple points on cleaved fresh surfaces and checked by multiple samples from the same batch using energy-dispersive x-ray spectroscopy in a JEOL LSM-6500 scanning electron microscope. The actual stoichiometry is Co$_{0.22(1)}$TaS$_{1.88(2)}$ with Ta fixed to 1, which is referred to as 2H-Co$_{0.22}$TaS$_2$ throughout this paper since $\sim 5\%$ sulfur deficiency is normally observed in 2H-TaS$_2$. X-ray diffraction (XRD) data were acquired on a Rigaku Miniflex powder diffractometer with Cu $K_{\alpha}$ ($\lambda=0.15418$ nm) at room temperature. X-ray absorption spectroscopy was measured at 8-ID beamline of the National Synchrotron Light Source II (NSLS II) at Brookhaven National Laboratory (BNL) in the fluorescence mode. The x-ray absorption near edge structure (XANES) and extended X-ray absorption fine structure (EXAFS) spectra were processed using the Athena software package. The EXAFS signal, $\chi(k)$, was weighed by $k^2$ to emphasize the high-energy oscillation and then Fourier-transformed in $k$ range from 2 to 10 {\AA}$^{-1}$ to analyze the data in $R$ space. The magnetization was measured in quantum design MPMS-XL5 system. Isotherms were collected at an interval of 1 K around $T_c$. The electrical and thermal transport were measured in quantum design PPMS-9 system. The longitudinal and Hall resistivity were measured using standard four-probe method. In order to effectively eliminate the longitudinal resistivity contribution due to voltage probe misalignment, the Hall resistivity was obtained by the difference of transverse resistance measured at positive and negative fields, i.e., $\rho_{xy}(\mu_0H) = [\rho(+\mu_0H)-\rho(-\mu_0H)]/2$.

\section{RESULTS AND DISCUSSIONS}

\begin{figure*}
\centerline{\includegraphics[scale=1]{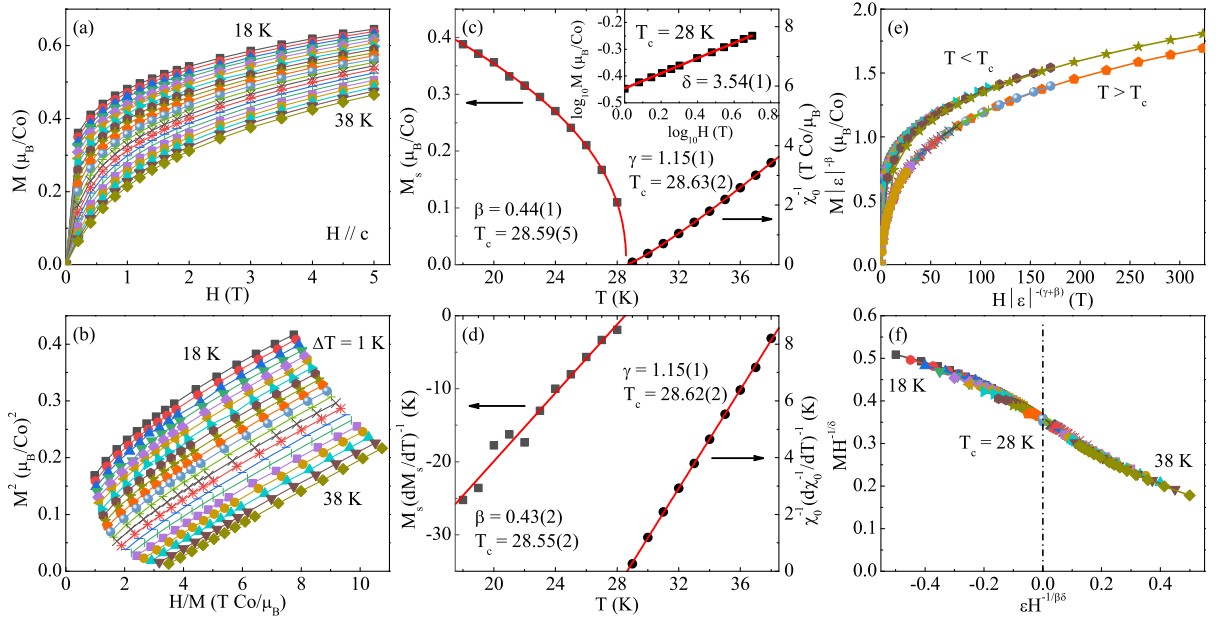}}
\caption{(Color online) (a) Typical initial isothermals measured with $\mathbf{H} \parallel \mathbf{c}$-axis from 18 to 38 K with a temperature step of 1 K. (b) The Arrott plots of $M^2$ vs $H/M$. (c) Temperature-dependent spontaneous magnetization $M_s$ (left) and inverse initial susceptibility $\chi_0^{-1}$ (right) with solid fitting curves for 2H-Co$_{0.22}$TaS$_2$. Inset shows the M(H) curve in log$_{10}$-log$_{10}$ scale collected at 28 K with linear fitting curve. (d) The Kouvel-Fisher plots of $M_s(dM_s/dT)^{-1}$ (left) and $\chi_0^{-1}(d\chi_0^{-1}/dT)^{-1}$ (right) with solid fitting curves. (e) Scaling plots of renormalized $m$ vs $h$ with $m\equiv\varepsilon^{-\beta}M(H,\varepsilon)$ and $h\equiv\varepsilon^{-(\beta+\gamma)}H$. (f) Rescaling plots of $M(H)$ isothermals by $MH^{-1/\delta}$ vs $\varepsilon H^{-1/(\beta\delta)}$.}
\label{XRD}
\end{figure*}

Figure 1(a,b) shows the crystal structure and the XRD 2theta scans for 2H-Co$_{0.22}$TaS$_2$. The sharp peaks can be indexed with (00l) planes [Fig. 1(b)], indicating that the plate surface of crystal is normal to the $\mathbf{c}$-axis. The main powder XRD pattern can be fitted with the $P6_322$ space group using Le Bail analysis, resulting in the lattice parameters $a$ = 5.72(2) {\AA} and $c$ = 11.89(2) {\AA} [inset in Fig. 1(b)]. It adopts the same space group with smaller lattice parameters when compared with 2H-Co$_{0.33}$TaS$_2$ ($a$ = 5.74 {\AA} and $c$ = 11.93 {\AA}) \cite{Parkin1,VanLaar}. Figure 1(c) exhibits the normalized Co $K$-edge XANES. The threshold energy $E_0$ = 7717 eV obtained from the peak of derivative curve [inset in Fig. 1(c)] indicates the Co$^{2+}$ state. Figure 1(d) exhibits the Fourier transform magnitudes of EXAFS. In a single-scattering approximation, the EXAFS could be described by \cite{Prins}:
\begin{align*}
\chi(k) = \sum_i\frac{N_iS_0^2}{kR_i^2}f_i(k,R_i)e^{-\frac{2R_i}{\lambda}}e^{-2k^2\sigma_i^2}sin[2kR_i+\delta_i(k)],
\end{align*}
where $N_i$ is the number of neighbouring atoms at a distance $R_i$ from the photoabsorbing atom. $S_0^2$ is the passive electrons reduction factor, $f_i(k, R_i)$ is the backscattering amplitude, $\lambda$ is the photoelectron mean free path, $\delta_i$ is the phase shift, and $\sigma_i^2$ is the correlated Debye-Waller factor measuring the mean square relative displacement of the photoabsorber-backscatter pairs. The main peak corresponds well to the six nearest-neighbor Co-S [2.37(1) {\AA}] and then two next-nearest Co-Ta [3.01(7) {\AA}] in the Fourier transform magnitudes of EXAFS [Fig. 1(d)], extracted from the model fit with fixed coordination number. The peaks above 3.5 {\AA} are due to longer Co-S [4.07(5) {\AA}] and Co-Ta [4.45(5) {\AA}] distances, and the multiple scattering behavior. Local crystal structure environment of Co atom indicates that Co is well intercalated \cite{Parkin1,VanLaar} in the vdW gap of 2H-TaS$_2$ crystal at the interstitial position as described in \cite{VanLaar} for 2H-Co$_{0.33}$TaS$_2$; somewhat smaller average lattice parameters are likely due to smaller Co intercalation.

\begin{table*}
\caption{\label{tab1}Comparison of critical exponents of 2H-Co$_{0.22}$TaS$_2$ and 2H-Fe$_{0.26}$TaS$_2$ with different theoretical models. The MAP, KFP, and CI represent the modified Arrott plot, the Kouvel-Fisher plot, and the critical isotherm, respectively.}
\begin{ruledtabular}
\begin{tabular}{lllllll}
   & Reference & Technique & $T_c$ & $\beta$ & $\gamma$ & $\delta$ \\
  \hline
  2H-Co$_{0.22}$TaS$_2$ & This work & MAP & 28.61(7) & 0.44(1) & 1.15(1) & 3.61(4)\\
  &  & KFP & 28.59(6) & 0.43(2) & 1.15(1) & 3.67(8) \\
  &  & CI  &  28  &   &   & 3.54(1) \\
  2H-Fe$_{0.26}$TaS$_2$ & \cite{Zhang} & MAP & 100.67(4) & 0.460(4) & 1.216(11) & 3.643(1)\\
  &  & KFP & 100.69(5) & 0.459(6) & 1.205(11) & 3.625(10) \\
  &  & CI  & 100.7 &   &   & 3.69(1) \\
  2D Ising & \cite{Widom} & Theory & & 0.125 & 1.75 & 15.0 \\
  Mean field & \cite{Kouvel} & Theory & & 0.5 & 1.0 & 3.0 \\
  3D Heisenberg & \cite{Kouvel} & Theory & & 0.365 & 1.386 & 4.8 \\
  3D Ising & \cite{Kouvel} & Theory & & 0.325 & 1.24 & 4.82 \\
  3D XY & \cite{Phan} & Theory & & 0.345 & 1.316 & 4.81 \\
  Tricritical mean field & \cite{Fisher1972} & Theory & & 0.25 & 1.0 & 5.0
\end{tabular}
\end{ruledtabular}
\end{table*}

Figure 1(c,d) exhibits the temperature dependence of dc magnetization measured in various fields applied in $\mathbf{ab}$-plane and along $\mathbf{c}$-axis with zero-field cooling (ZFC) and field cooling (FC) modes, respectively. A sharp upturn in $M(T)$ was observed for both field directions when temperature decreases, suggesting a PM-FM transition. The $M(T)$ for $\mathbf{H} \parallel \mathbf{c}$ are larger than those for $\mathbf{H} \parallel \mathbf{ab}$, indicating an uniaxial anisotropy in 2H-Co$_{0.22}$TaS$_2$. The difference between ZFC and FC curves is due to magnetic domain creep effect. The $1/\chi(T)$ = $H/M(T)$ taken at 1 kOe from 100 to 300 K can be well fitted by the Curie-Weiss law $\chi = \chi_0 + C/(T-\theta)$ [insets in Fig. 1(c,d)], where $\chi_0$ is a temperature-independent term, $C$ and $\theta$ are the Curie-Weiss constant and Weiss temperature, respectively. The Weiss temperature is $\theta_c$ = 40(2) K for $\mathbf{H} \parallel \mathbf{c}$ and $\theta_{ab}$ = -5(1) K for $\mathbf{H} \parallel \mathbf{ab}$, respectively, indicating dominance of FM exchange interactions along $\mathbf{c}$-axis and AFM in $\mathbf{ab}$-plane in 2H-Co$_{0.22}$TaS$_2$. The derived effective moment is 2.3(1) $\mu_\textrm{B}$/Co for $\mathbf{H} \parallel \mathbf{ab}$ and 2.7(1) $\mu_\textrm{B}$/Co for $\mathbf{H} \parallel \mathbf{c}$, respectively, which is smaller than the spin-only moment of 3.87 $\mu_\textrm{B}$ for Co$^{2+}$. The deviation between 2.3(1) and 2.7(1) $\mu_\textrm{B}$/Co may be due to the anisotropic $g$ factor
with unquenched orbital angular moment. Future experiments such as neutron scattering or electron spin resonance would be helpful to clarify this difference. Figure 1(e,f) presents the field-dependent magnetization at various temperatures. A large coercivity of 14.3 kOe was observed at 2 K when $\mathbf{H} \parallel \mathbf{c}$, which decreases to 0.7 kOe at 10 K and disappears at 50 K. The saturation moment is also relatively small ($<$ 0.8 $\mu_\textrm{B}$/Co), suggesting that both localized and delocalized Co electrons coexist in 2H-Co$_{0.22}$TaS$_2$. Only linear dependent $M(H)$ was observed when $\mathbf{H} \parallel \mathbf{ab}$, unambiguously confirming a large magnetocrystalline anisotropy in 2H-Co$_{0.22}$TaS$_2$. This is the first time to achieve FM order with uniaxial anisotropy in Co-intercalated 2H-TaS$_2$ crystals. The FM coupling between intercalated-Co moments in 2H-TaS$_2$ layer may arise from an indirect spin exchange interaction, namely, the Ruderman-Kittel-Kasuya-Yosida (RKKY) interaction \cite{Aristov,Yosida,Fukuma}. The high carrier density in 2H-TaS$_2$ host also favors the spin exchange interactions between isolated Co atoms through the RKKY mechanism. Similar mechanism was proposed for magnetic order in the related materials Cr$_{0.33}$TaS$_2$ and V$_{0.25}$VS$_2$ \cite{YamasakiY,NiuJ}.

To obtain a precise $T_c$ and the mechanism of PM-FM transition in 2H-Co$_{0.22}$TaS$_2$, we measured dense magnetization isotherms with $\mathbf{H} \parallel \mathbf{c}$ from 18 to 38 K [Fig. 2(a)]. The critical behavior gives insight into the nature of magnetic interactions, correlation length, spin dimensionality, and the spatial decay of correlation function at criticality \cite{Fisher0,Stanley0}. The Arrott plot ($M^2$ vs $H/M$) [Fig. 2(b)] shows quasi-straight and -parallel lines in high field region \cite{Arrott1}. To obtain the critical exponents, the isotherms were then reanalyzed with the Arrott-Noakes equation of state \cite{Arrott2},
\begin{equation}
(H/M)^{1/\gamma} = a\varepsilon+bM^{1/\beta},
\end{equation}
where $\varepsilon = (T-T_c)/T_c$ is the reduced temperature, and $a$ and $b$ are constants. The $\beta$ and $\gamma$ are associated with the spontaneous magnetization $M_s$ below $T_c$ and the inverse initial susceptibility $\chi_0^{-1}$ above $T_c$, respectively \cite{Stanley,Fisher,Lin}:
\begin{equation}
M_s (T) = M_0(-\varepsilon)^\beta, \varepsilon < 0, T < T_c,
\end{equation}
\begin{equation}
\chi_0^{-1} (T) = (h_0/m_0)\varepsilon^\gamma, \varepsilon > 0, T > T_c,
\end{equation}
\begin{equation}
M = DH^{1/\delta}, T = T_c,
\end{equation}
where $\delta$ is another critical exponent associated with the M(H) at $T_c$. The $M_0$, $h_0/m_0$ and $D$ are the critical amplitudes. A rigorous iterative method was adopted with an initial mean field model \cite{Kellner,Pramanik}.

Figure 2(c) exhibits the temperature dependence of final $M_s(T)$ and $\chi_0^{-1}(T)$ with power law fitting that gives $\beta = 0.44(1)$ with $T_c = 28.59(5)$ K, and $\gamma = 1.15(1)$ with $T_c = 26.63(2)$ K. The third exponent $\delta$ = 3.54(1) is determined by a linear fit of $M(H)$ at 28 K taking into account that $M = DH^{1/\delta}$ at $T_c$ [inset in Fig. 2(c)], which is very close to the value of 3.61(4) calculated from the Widom scaling relation $\delta = 1+\gamma/\beta$ \cite{Widom}. In the Kouvel-Fisher (KF) relation \cite{Kouvel}:
\begin{equation}
M_s(T)[dM_s(T)/dT]^{-1} = (T-T_c)/\beta,
\end{equation}
\begin{equation}
\chi_0^{-1}(T)[d\chi_0^{-1}(T)/dT]^{-1} = (T-T_c)/\gamma.
\end{equation}
Linear fittings to the plots of $M_s(T)[dM_s(T)/dT]^{-1}$ vs $T$ and $\chi_0^{-1}(T)[d\chi_0^{-1}(T)/dT]^{-1}$ vs $T$ yield $\beta = 0.43(2)$ with $T_c = 28.55(2)$ K, and $\gamma = 1.15(1)$ with $T_c = 28.62(2)$ K, as shown in Fig. 2(d), very close to the values obtained from the modified Arrott plot. Scaling analysis was then used to double-check the self-consistency and reliability of the obtained $\beta$, $\gamma$, $\delta$, and $T_c$.

\begin{figure*}
\centerline{\includegraphics[scale=1]{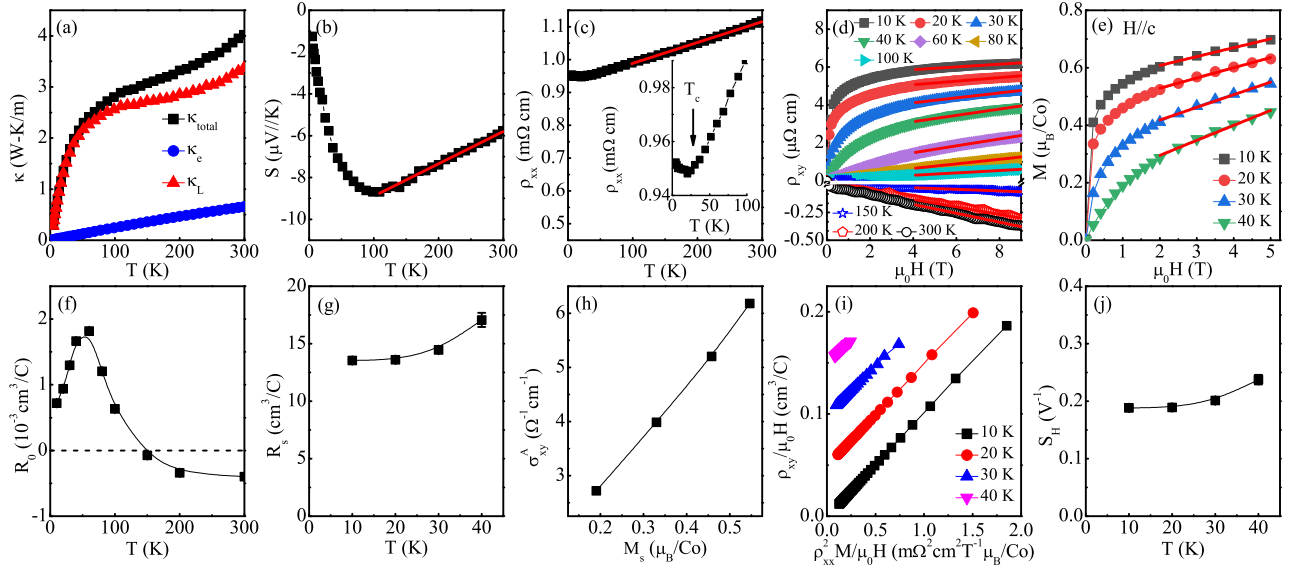}}
\caption{(Color online) Temperature dependence of in-plane (a) thermal conductivity $\kappa(T)$, (b) thermopower $S(T)$, and (c) resistivity $\rho_{xx}(T)$ in zero field for 2H-Co$_{0.22}$TaS$_2$ single crystal. Out-of-plane field dependence of (d) Hall resistivity $\rho_{xy}(\mu_0H)$ and (e) initial magnetization $M(\mu_0H)$ for 2H-Co$_{0.22}$TaS$_2$ at indicated temperatures. Temperature dependence of (f) ordinary Hall coefficient $R_0$ and (g) anomalous Hall coefficient $R_s$, fitted from the $\rho_{xy}(\mu_0H)$ curves using $\rho_{xy} = R_0\mu_0H + R_sM$. (h) Anomalous Hall conductivity $\sigma_{xy}^A$ vs $M_s$ with $\sigma_{xy}^A \approx \rho_{xy}^A / \rho_{xx}^2$. (i) The $\rho_{xy}/\mu_0H$ vs $\rho_{xx}^2M/\mu_0H$ curves at indicated temperatures with subsequent offset of 0.05 cm$^3$ C$^{-1}$. (j) Temperature dependence of coefficient $S_H=\mu_0R_s/\rho_{xx}^2$.}
\label{MTH}
\end{figure*}

The magnetic equation of state in the critical region is expressed as \cite{Stanley}:
\begin{equation}
M(H,\varepsilon) = \varepsilon^\beta f_\pm(H/\varepsilon^{\beta+\gamma}),
\end{equation}
where $f_+$ for $T>T_c$ and $f_-$ for $T<T_c$, respectively, are the regular functions. Eq. (7) can be further written in terms of scaled magnetization $m\equiv\varepsilon^{-\beta}M(H,\varepsilon)$ and scaled field $h\equiv\varepsilon^{-(\beta+\gamma)}H$ as $m = f_\pm(h)$. This suggests that for true scaling relations and the right choice of critical exponents, scaled $m$ and $h$ will fall on universal curves above $T_c$ and below $T_c$, respectively. The scaled $m$ vs $h$ curves are obviously seperated into two branches below and above $T_c$, respectively, as shown in [Fig. 2(e)]. This can also be checked by another form of scaling equation of state \cite{Stanley},
\begin{equation}
\frac{H}{M^\delta} = k\left(\frac{\varepsilon}{H^{1/\beta}}\right),
\end{equation}
where $k(x)$ is the scaling function. According to Eq. (8), the $MH^{-1/\delta}$ vs $\varepsilon H^{-1/(\beta\delta)}$ curves should be scaled into one universal curve \cite{Phan}, which is indeed seen in Fig. 2(f). The well-rescaled curves ensure that the obtained critical exponents and $T_c$ are reliable and intrinsic.

The critical exponents of 2H-Co$_{0.22}$TaS$_2$ obtained using various methods as well as different theoretical models are summarized in Table I. Taroni \emph{et al.} pointed out that the value of $\beta$ for a 2D magnet should be within a window $0.1 \leq \beta \leq 0.25$ \cite{Taroni}, indicating a clear 3D critical behavior in 2H-Co$_{0.22}$TaS$_2$. Moreover, the critical exponents do not belong to any single universality class. The $\beta$ lies between 3D Heisenberg and mean field model but closer to the latter, indicating the exchange interaction is of long-range type in 2H-Co$_{0.22}$TaS$_2$. However, the $\gamma$ lies between 3D Ising and mean field model, which is attributed to strong uniaxial anisotropy in 2H-Co$_{0.22}$TaS$_2$. The mechanism of FM coupling in 2H-Co$_{0.22}$TaS$_2$ is the RKKY interaction in which the local spins of intercalated Co ions align ferromagnetically through the itinerant Ta 5d electrons. Compared with 2H-Fe$_{0.26}$TaS$_2$, the slightly smaller values of $\beta$ and $\gamma$ suggest a similar critical behavior in 2H-Co$_{0.22}$TaS$_2$; i.e., a localized 3D Heisenberg type coupled with long range magnetic interactions \cite{Zhang}.

It is also of interest to estimate the range of interaction in this material. In renormalization group theory analysis the interaction decays with distance $r$ as $J(r) \approx r^{-(3+\sigma)}$ for long-range exchange, where $\sigma$ is the range of exchange interaction \cite{Fisher1972}. The susceptibility exponent $\gamma$ is:
\begin{multline}
\gamma = 1+\frac{4}{d}\left(\frac{n+2}{n+8}\right)\Delta\sigma+\frac{8(n+2)(n-4)}{d^2(n+8)^2}\\\times\left[1+\frac{2G(\frac{d}{2})(7n+20)}{(n-4)(n+8)}\right]\Delta\sigma^2,
\end{multline}
where $\Delta\sigma = (\sigma-\frac{d}{2})$, $G(\frac{d}{2})=3-\frac{1}{4}(\frac{d}{2})^2$, and $n$ is the spin dimensionality \cite{Fischer}. When $\sigma > 2$, it indicates a short-range interaction, while $\sigma < 2$ implies a long-range type of interaction, and mean-field model should apply when $\sigma < 3/2$. According to Eq. (9), it is found that \{$d:n$\} = \{3 : 3\} and $\sigma = 1.69$ give the critical exponents $\beta = 0.44$, $\gamma = 1.14$, and $\delta = 3.59$ [$\nu = \gamma/\sigma$, $\alpha = 2-\nu d$, $\beta = (2-\alpha-\gamma)/2$, and $\delta = 1 + \gamma/\beta$], which are mostly close to our experimentally observed values. The obtained $\sigma = 1.69$ is slightly smaller than that of 1.71 for 2H-Fe$_{0.26}$TaS$_2$, in line with the comparison of $\beta$ and $\gamma$. It confirms that the spin interaction in 2H-Co$_{0.22}$TaS$_2$ is of 3D Heisenberg (\{$d:n$\} = \{3 : 3\}) type coupled with long range magnetic interactions and that the exchange decays with distance as $J(r)\approx r^{-4.69}$; magnetic anisotropy probably originates from the unquenched orbital moment and different crystallographic environment for Co atoms in the vdW gap plane and along the $\mathbf{c}$-axis of crystal.

Having established magnetic properties and the nature of critical behavior, we proceed to investigate effects of Co intercalation on electronic transport properties. Figure 3(a-c) shows the temperature-dependent thermal conductivity $\kappa(T)$, thermopower $S(T)$, and in-plane resistivity $\rho_{xx}(T)$ of 2H-Co$_{0.22}$TaS$_2$ single crystal. It is obvious that Co intercalation removes charge density wave transition at 78 K \cite{Harper}. Generally, $\kappa_{total} = \kappa_e + \kappa_L$, consists of the electronic part $\kappa_e$ and the phonon term $\kappa_L$. The $\kappa_L$ can be obtained by subtracting the $\kappa_e$ part calculated from the Wiedemann-Franz law $\kappa_e/T = L_0/\rho$, where $L_0$ = 2.45 $\times$ 10$^{-8}$ W $\Omega$ K$^{-2}$ \cite{WH}. The calculated $\kappa_e$ and $\kappa_L$ are also plotted in Fig. 3(a), indicating that the $\kappa_L$ dominates. Above 100 K, the linear temperature dependence of $S(T)$ and $\rho(T)$ is observed since the magnetic scattering is independent of temperature in the paramagnetic regime. The negative values of $S(T)$ in the whole temperature range indicates that electron-type carriers dominate; it changes the slope below 100 K [Fig. 3(b)]. There is no obvious anomaly around $T_c$ in $S(T)$, probably due to the fact that ordered moment is small [Figs. 1(g,h)] since the Co does not develop full moment due to hybridization with conduction electron band. This ia also the reason for the relatively weak upturn in $\rho(T)$ below $T_c$ [inset in Fig. 3(c)].

Figure 3(d) shows the field dependence of Hall resistivity $\rho_{xy}(\mu_0H)$ of 2H-Co$_{0.22}$TaS$_2$ measured at various temperatures. Above 100 K, the negative slope of $\rho_{xy}(\mu_0H)$ indicates the dominance of electron-type carries, in line with the $S(T)$ analysis, which can be accounted for by the ordinary Hall coefficient. With decreasing temperature, the Hall coefficient changes sign from negative to positive, in line with the slope change of $S(T)$. It gradually deviates from the linear field-dependence and varies non-linearly with the applied magnetic field around $T_c$. With approaching $T_c$, the tendency is similar with the $M(\mu_0H)$ curves [Fig. 3(e)], indicating that there is AHE in 2H-Co$_{0.22}$TaS$_2$. The AHE originates from the spontaneous polarization induced by non-zero net magnetization due to significant hybridization between Co-3d and Ta-5d bands in this system.

Generally, the Hall resistivity in ferromagnets is made up of two parts \cite{Wang,Yan, WangY,Onoda2008},
\begin{equation}
\rho_{xy} = \rho_{xy}^O + \rho_{xy}^A = R_0\mu_0H + R_sM,
\end{equation}
where $\rho_{xy}^O$ and $\rho_{xy}^A$ are the ordinary and anomalous Hall resistivity, respectively. The $R_0$ is the ordinary Hall coefficient and the $R_s$ is the anomalous Hall coefficient. With linear fits of $\rho_{xy}(\mu_0H)$ in high field, the slope and intercept correspond to $R_0$ and $\rho_{xy}^A$, respectively. The $R_s$ can be obtained from $\rho_{xy}^A = R_sM_s$ with $M_s$ taken from linear fits of $M(\mu_0H)$ curves in high field. The temperature dependence of derived $R_0$ and $R_s$ is plotted in Fig. 3(f,g), in which the value of $R_s$ is about ten times larger than that of $R_0$ below $T_c$. The $R_0$ shows a remarkable temperature dependence, which changes sign at high temperature and reaches a maximum at low temperature [Fig. 3(f)]. Combined with the slope change of $S(T)$ below 100 K, this points to a possible multi-band behavior.

Figure 3(h) exhibits the anomalous Hall conductivity (AHC) $\sigma_{xy}^A$ ($\approx$ $\rho_{xy}^A / \rho_{xx}^2$), which is proportional to $M_s$, usually expected by the Kaplus-Luttinger (KL) theory \cite{KL}. The value of anomalous Hall conductivity is found to be very small compared to usual FM conductors since the density of ferromagnetic Co is low. The intrinsic KL contribution of $\sigma_{xy,in}^A$ is of the order of $e^2/(hd)$, where $e$ is the electronic charge, $h$ is the Plank constant, and $d$ is the lattice parameter \cite{Onoda2006}. Taking $d \approx V^{1/3} \sim 7$ {\AA}, $\sigma_{xy,in}^A$ is estimated to be 550 ($\Omega$ cm)$^{-1}$, which is much larger than the obtained values in Fig. 3(h); this precludes the possibility that KL mechanism is observed in experiment. The extrinsic side-jump contribution of $\sigma_{xy,sj}^A$ is usually of the order of $e^2/(hd)(\varepsilon_{SO}/E_F)$, where $\varepsilon_{SO}$ and $E_F$ is the spin-orbital interaction energy and the Fermi energy, respectively \cite{Nozieres}. The value of $\varepsilon_{SO}/E_F$ is generally less than $10^{-2}$ for metallic ferromagnets, which points to the same order of magnitude of the obtained $\sigma_{xy}^A$. Furthermore, the extrinsic side-jump mechanism, where the potential field induced by impurities contributes to the anomalous group velocity, follows a scaling behavior of $\rho_{xy}^A = \beta\rho_{xx}^2$, the same with the intrinsic KL mechanism, however, the skew-scattering mechanism describes asymmetric scattering induced by impurities or defects and contributes to the AHE with a scaling behavior of $\rho_{xy}^A = \beta\rho_{xx}$. As we can see, the scaling plots of the modified anomalous Hall resistivity $\rho_{xy}/\mu_0H_{eff}$ and longitudinal resistivity $\rho_{xx}^2M/\mu_0H_{eff}$, with subsequent offset by 0.05 cm$^3$ C$^{-1}$, shows a good linear behavior [Fig. 3(i)]. This linear scaling behavior indicates that the AHE in 2H-Co$_{0.22}$TaS$_2$ is probably dominated by the extrinsic side-jump mechanism rather than extrinsic skew-scattering and the intrinsic KL theory. Then Eq. (10) can be written as
\begin{equation}
\rho_{xy} = R_0\mu_0H + S_H\rho_{xx}^2M,
\end{equation}
where $S_H$ is a material-specific scale factor \cite{Lee1,Fri}. The coefficient $S_H = \mu_0R_s/\rho_{xx}^2 = \sigma_{xy}^A/M_s$ [Fig. 3(j)] is found to be weakly temperature-dependent and is comparable with those in traditional itinerant ferromagnets, such as Fe and Ni ($S_H \sim 0.01 - 0.2$ V$^{-1}$) \cite{Dheer,Jan}.

\section{CONCLUSIONS}

In summary, we first synthesized single crystals of ferromagnetic 2H-Co$_{x}$TaS$_2$ by intercalating $x \sim$ 0.22 Co in 2H-TaS$_2$ vdW gap. We have also systematically studied the second-order PM-FM transition in 2H-Co$_{0.22}$TaS$_2$ with $T_c$ $\sim$ 28 K. Critical exponents $\beta$, $\gamma$, $\delta$, and $T_c$ derived via various methods match reasonably well and follow the scaling equation. The detailed analysis indicates that the spin interaction in 2H-Co$_{0.22}$TaS$_2$ is of 3D Heisenberg-type coupled with long-range exchange interaction decaying with distance as $J(r)\approx r^{-4.69}$. The origin of the AHE below $T_c$ in 2H-Co$_{0.22}$TaS$_2$ crystal is probably dominated by the extrinsic side-jump mechanism rather than extrinsic skew-scattering and intrinsic KL mechanisms. Since few-layer graphene/2H-TaS$_2$ heterostructures preserve 2D Dirac states with robust spin-helical structure of interest for spin-logic circuits \cite{LiLijun}, the possibility of integration of robust ferromagnetism is of interest for spintronic and calls for nanofabrication of graphene/2H-Co$_{0.22}$TaS$_2$ heterostructures and devices.

\section*{Acknowledgements}

Work at BNL is supported by the Office of Basic Energy Sciences, Materials Sciences and Engineering Division, U.S. Department of Energy (DOE) under Contract No. DE-SC0012704. This research used the 8-ID beamline of the NSLS II, a U.S. DOE Office of Science User Facility operated for the DOE Office of Science by BNL under Contract No. DE-SC0012704.\\

$^{*}$Present address: Los Alamos National Laboratory, MS K764, Los Alamos NM 87545, USA\\

$^{\dag}$Present address: ALBA Synchrotron Light Source, Cerdanyola del Valles, E-08290 Barcelona, Spain.

\end{document}